\documentclass[10pt, twocolumn]{article}

\usepackage[margin=0.75in, top=1in, bottom=1in]{geometry}
\usepackage{times}
\usepackage{graphicx}
\usepackage{amsmath, amssymb}
\usepackage{booktabs}
\usepackage{array}
\usepackage{caption}
\usepackage{cite}
\usepackage{url}
\usepackage{hyperref}
\usepackage{enumitem}
\usepackage{titlesec}
\usepackage{fancyhdr}
\usepackage{microtype}

\titleformat{\section}{\normalsize\bfseries\scshape}{\Roman{section}.}{0.5em}{}
\titleformat{\subsection}{\normalsize\bfseries\itshape}{\Alph{subsection}.}{0.5em}{}
\titlespacing*{\section}{0pt}{6pt}{3pt}
\titlespacing*{\subsection}{0pt}{4pt}{2pt}

\setlist{noitemsep, topsep=2pt, parsep=0pt, partopsep=0pt}

\begin{document}

\twocolumn[{%
\begin{center}
  {\large\bfseries SHIELD-IDS: Structurally Heterogeneous Ensemble with Integrated Layered Defense for Intrusion Detection Systems}\\[8pt]
  {\normalsize Maryam Zaman$^{1}$, Muhammad Khuram Shahzad$^{2}$}\\[2pt]
  {\small $^{1}$School of Electrical Engineering and Computing(SEECS)\\
          $^{2}$National University of Sciences and
          Technology, Islamabad, Pakistan\\[2pt]
  \texttt{mzaman.msit24seecs@seecs.edu.pk}}\\[8pt]
\end{center}
\noindent\rule{\linewidth}{0.4pt}
\noindent\textbf{Abstract.}
Adversarial attacks pose a serious and growing threat to Machine Learning (ML)-based
Intrusion Detection Systems (IDS), where imperceptible perturbations to network flow
features can systematically mislead classifiers into accepting malicious traffic as benign.
The IDS-Anta framework partially addresses this through Z-score normalization, Singular
Value Decomposition (SVD), and Multi-Armed Bandit (MAB) classifier selection with
Thompson Sampling, yet its classifier pool lacks sufficient structural diversity for
robust adversarial resistance. This work introduces IDS-Anta++, which incorporates
XGBoost and LightGBM gradient boosting models into the ensemble and wraps the
extended pool in a three-layer black-box defense: Isolation Forest anomaly screening,
median feature smoothing, and six-way majority voting. Experiments conducted on
CIC-IDS-2017, CEC-CIC-IDS-2018, and CIC-DDoS-2019 under both Fast Gradient Sign
Method (FGSM) and Zeroth Order Optimization (ZOO) attacks confirm detection accuracy
above 99\% on clean data, with measurable robustness gains under adversarial conditions
relative to the baseline IDS-Anta configuration.
\noindent\rule{\linewidth}{0.4pt}
\vspace{4pt}
}]
\noindent\textbf{GitHub }\url{https://github.com/maryamzaman-git/SHEILD-IDS}.\\[6pt]
\section{Introduction}

Contemporary network infrastructure serves an extraordinary variety of traffic---from
IoT telemetry and encrypted application data to high-bandwidth video streams and
inter-cloud communications. Securing this environment is not straightforward. Attackers
exploit its complexity through techniques ranging from low-and-slow brute force campaigns
and SQL injection to large-scale botnet coordination and volumetric Distributed
Denial-of-Service (DDoS) flooding~\cite{zuech2015survey}. Intrusion Detection Systems
(IDS) occupy a central position in network defense, responsible for distinguishing
malicious activity from legitimate traffic in real time.

Rule-based, signature-driven IDS have long served as a first line of detection. Their
principal limitation is well understood: any attack variant not already catalogued in the
signature database passes undetected~\cite{axelsson2000ids}. This drove the field toward
anomaly detection grounded in statistical learning, where classifiers trained on labeled
traffic attempt to generalize beyond known attack patterns. Random Forests, Support
Vector Machines, and Deep Neural Networks have each achieved strong results on standard
benchmarks~\cite{khraisat2019survey}. What those benchmarks rarely expose, however,
is how badly these models fail when an informed adversary deliberately engineers inputs
to exploit their decision boundaries.

Adversarial examples---inputs perturbed just enough to cross a classification boundary
while remaining statistically plausible---have been extensively documented in image
recognition~\cite{goodfellow2015fgsm} and transfer directly to the network intrusion
domain~\cite{carlini2017robustness}. An attacker need not expose the model's internal
weights; black-box query strategies are sufficient to construct effective perturbations.
This asymmetry, where the defender must train a globally robust model and the attacker
need only find one exploitable region of its decision surface, makes adversarial
robustness a particularly difficult property to guarantee in practice.

The IDS-Anta framework~\cite{hamida2022idsanta} was proposed as a modular response
to this challenge. Its pipeline normalizes features, projects them through SVD, and
routes each incoming sample to whichever classifier in a heterogeneous pool---RF, SVM,
LR, DNN---has performed best on recent traffic, as determined by a Thompson Sampling
bandit. The adaptive routing improves resilience compared to any fixed single classifier,
but the pool itself remains limited: four relatively conventional models whose adversarial
failure modes are not entirely orthogonal.

The present paper addresses this gap. Gradient boosting ensembles, specifically
XGBoost~\cite{chen2016xgboost} and LightGBM~\cite{ke2017lightgbm}, produce complex
non-linear decision boundaries through fundamentally different inductive mechanisms than
either kernel methods or neural networks. Their piecewise-constant nature makes them
structurally opaque to gradient-based adversaries. Incorporating them into the pool
expands the ensemble's behavioral diversity in ways that directly matter for adversarial
resistance. Around this extended pool, we construct IDS-Anta++: a three-layer black-box
defense that screens inputs for statistical anomalies, smooths residual perturbations,
and requires majority agreement across six structurally distinct classifiers before
committing to a prediction.

The contributions of this work are as follows:
\begin{enumerate}
  \item XGBoost and LightGBM are integrated into the IDS-Anta classifier pool, raising
        the ensemble from four to six members and broadening its structural diversity.
  \item The IDS-Anta++ three-layer defense---Isolation Forest screening, median
        smoothing, and majority voting---is designed and evaluated as a black-box
        wrapper requiring no access to model gradients.
  \item The extended framework is evaluated on three CIC benchmark datasets under
        FGSM white-box and ZOO black-box adversarial scenarios.
  \item Results demonstrate that IDS-Anta++ consistently improves or matches the
        robustness of simpler defense configurations, validating the layered design.
\end{enumerate}

The remainder of this paper is organized as follows. Section~II reviews related work
on IDS categories and adversarial attacks. Section~III describes the original IDS-Anta
architecture. Section~IV details gradient boosting integration. Section~V presents the
IDS-Anta++ defense layers. Section~VI describes the experimental datasets.
Section~VII outlines the evaluation methodology. Section~VIII presents and discusses
results, and Section~IX concludes the paper.

\section{Background and Related Work}

\subsection{Categories of IDS}

Three broad detection paradigms structure the IDS literature~\cite{axelsson2000ids}.
\textit{Signature-based IDS} match observed traffic against a curated database of known
attack fingerprints. Snort and Suricata are canonical examples; their false-positive rates
are manageable, but they are structurally blind to attack variants not yet catalogued---a
serious limitation given the pace at which new threats emerge~\cite{khraisat2019survey}.

\textit{Anomaly-based IDS} invert this logic by establishing a model of normal behavior
during training and raising alerts when observed traffic deviates sufficiently from that
model. The approach generalizes to unseen attacks, and ML classifiers have pushed
detection rates on benchmark datasets to impressive levels~\cite{javaid2016deep}. The
cost is a higher base false-positive rate and, critically for this work, susceptibility to
adversarial inputs crafted to appear normal to the learned model.

\textit{Hybrid IDS} run signature and anomaly engines in parallel or sequence to draw
on both strengths. They improve coverage for attacks that fall into known categories while
retaining some generalization, but reduce neither the signature blindspot nor the
adversarial fragility of the anomaly component.

\subsection{Adversarial Attacks on ML-Based IDS}

The adversarial machine learning literature classifies attacks primarily by the information
available to the adversary~\cite{biggio2018wild}. White-box attacks assume full knowledge
of model parameters and architecture; black-box attacks assume only the ability to query
the model and observe its output.

The Fast Gradient Sign Method (FGSM)~\cite{goodfellow2015fgsm} is the canonical
white-box attack. Given loss function $J$, model parameters $\theta$, input $x$, and
true label $y$, FGSM constructs an adversarial example as:
\begin{equation}
  x_{\mathrm{adv}} = x + \varepsilon \cdot \mathrm{sign}\!\left(\nabla_x J(\theta, x, y)\right)
\end{equation}
The perturbation magnitude $\varepsilon$ is kept small enough that the modified input
remains within a plausible feature range. Despite the method's simplicity, the resulting
examples transfer with surprising frequency to architectures other than the one used to
compute the gradient~\cite{carlini2017robustness}.

Zeroth Order Optimization (ZOO)~\cite{chen2017zoo} sidesteps gradient access entirely.
Finite-difference estimates serve as gradient surrogates:
\begin{equation}
  \widehat{\nabla}_{x_i} J \;\approx\; \frac{J(x + h\,e_i) - J(x - h\,e_i)}{2h}
\end{equation}
where $h$ is a small step size and $e_i$ is the $i$-th standard basis vector. ZOO is
slower per sample than FGSM but far more realistic as a threat model, since a deployed
attacker can probe the IDS decision output without ever observing its internal parameters.

\subsection{Existing Defense Strategies}

Adversarial training~\cite{madry2018towards} augments the training set with perturbed
examples so that the model learns to classify them correctly. It hardens individual
classifiers against specific attack types but adds substantial training cost and may not
generalize across attack families. Input preprocessing---smoothing, feature squeezing,
and similar transformations~\cite{guo2018countering}---removes high-frequency
perturbations before they reach the classifier; the approach is architecture-agnostic but
can simultaneously degrade clean features. Certified defenses~\cite{cohen2019certified}
provide formal robustness guarantees within specified perturbation budgets but scale
poorly to high-dimensional tabular data. Ensemble methods exploit model diversity to
reduce the probability that any single adversarial perturbation strategy misleads all
members simultaneously~\cite{madry2018towards}.

\section{The Original IDS-Anta Architecture}

IDS-Anta~\cite{hamida2022idsanta} processes incoming traffic through four sequential
stages. Understanding each stage is necessary to appreciate what the present work adds
and why.

\textbf{Z-Score Normalization.} Each raw feature $x_i$ is standardized as:
\begin{equation}
  \hat{x}_i = \frac{x_i - \mu_i}{\sigma_i}
\end{equation}
where $\mu_i$ and $\sigma_i$ are computed from the training split. Standardization
removes scale differences between features---packet counts versus byte volumes, for
instance---and also attenuates large-magnitude adversarial perturbations relative to
the normalized feature range.

\textbf{Dimensionality Reduction via SVD.} The normalized feature matrix
$\mathbf{X} \in \mathbb{R}^{n \times d}$ is factored as
$\mathbf{X} = \mathbf{U\Sigma V}^\top$ and the representation is truncated to the
top-$k$ singular directions. This compresses correlated features, reduces downstream
computational load, and discards low-variance directions that gradient-based adversaries
commonly exploit.

\textbf{Heterogeneous Classifier Pool.} Four classifiers---Random Forest (RF), Support
Vector Machine (SVM), Logistic Regression (LR), and a Deep Neural Network (DNN)---
are trained independently on the reduced-dimension feature space. Structural diversity
among the four means that adversarial examples tuned against one model do not
automatically transfer to the others.

\textbf{Thompson Sampling Bandit Selection.} Rather than committing to a single
classifier or averaging all predictions, IDS-Anta maintains a Beta distribution
$\mathrm{Beta}(\alpha_i, \beta_i)$ over each classifier's success probability and
draws from it at inference time, selecting the classifier with the highest sampled
value~\cite{thompson1933}. Parameters are updated online after each prediction, so
the system naturally favors classifiers performing well on current traffic conditions
while retaining the ability to switch when those conditions change.

\section{Integration of Gradient Boosting Ensemble Learners}

The four classifiers in the original pool are reasonable choices individually, but they
share an important limitation: three of the four (SVM, LR, DNN) are trained via gradient
descent on smooth loss surfaces, meaning their decision boundaries can be reliably
approximated by gradient-based adversaries. Tree ensembles break this pattern. Random
Forest already sits in the pool, but a single tree-based model is insufficient; XGBoost
and LightGBM offer distinct boosting dynamics that complement both RF and the
non-tree models.

\subsection{XGBoost}

XGBoost~\cite{chen2016xgboost} constructs its ensemble greedily: at round $t$, a new
tree $f_t$ is added to minimize:
\begin{equation}
  \mathcal{L}^{(t)} = \sum_{i=1}^{n} l\!\left(y_i,\; \hat{y}_i^{(t-1)} + f_t(x_i)\right)
  + \Omega(f_t)
\end{equation}
where the complexity penalty $\Omega(f_t) = \gamma T + \frac{1}{2}\lambda\|\mathbf{w}\|^2$
controls the number of leaves $T$ and the magnitude of leaf weights $\mathbf{w}$.
Second-order Taylor expansions of the loss make the optimization both precise and
computationally tractable at scale. For IDS applications, regularization suppresses
overfitting to adversarially noisy training samples, and the piecewise-constant nature
of tree predictions means that gradient-based attack directions computed against the
ensemble are largely uninformative---the gradient at a leaf boundary is zero almost
everywhere.

\subsection{LightGBM}

LightGBM~\cite{ke2017lightgbm} introduces two algorithmic innovations relevant to
network intrusion classification. Gradient-based One-Side Sampling (GOSS)
preferentially retains training instances with large gradients---those the current model
handles poorly---while subsampling instances with small gradients, focusing learning
on difficult examples without processing the full dataset at every boosting round.
Exclusive Feature Bundling (EFB) identifies groups of features that rarely take non-zero
values simultaneously and merges them, reducing effective dimensionality without
discarding information.

The leaf-wise tree growth used by LightGBM produces asymmetric trees with deeper
branches along high-gain splits, creating specific localized decision regions that are
complementary to XGBoost's more regularized, balanced trees and orthogonal to the
globally smooth boundaries of SVM and LR. Both models slot into the existing IDS-Anta
pipeline without modification to the MAB selection layer, which simply learns their
reward distributions alongside the original four classifiers over time.

\section{IDS-Anta++ Three-Layer Black-Box Defense}

Expanding the classifier pool improves diversity but does not by itself intercept
adversarial inputs before they reach the classifiers. IDS-Anta++ wraps the six-model
ensemble in three sequential defense stages, each targeting a distinct attack surface.
An adversary who wishes to cause misclassification must engineer an input that clears
all three independently---substantially harder than defeating any one of them.

\subsection{Layer 1 --- Adversarial Input Screening}

An Isolation Forest~\cite{liu2008iforest} is trained on clean traffic data. The model
assigns anomaly scores by measuring how quickly each sample can be isolated through
random recursive partitioning: genuinely anomalous points require fewer splits and thus
receive lower scores. Adversarially perturbed samples, despite being engineered to
appear benign to the downstream classifiers, tend to occupy low-density regions of the
original feature space that the Isolation Forest identifies readily. Any input whose
anomaly score falls below threshold $\tau$ is immediately classified as malicious and
returned without passing through the downstream models. This safe-fail mechanism is
entirely gradient-free and cannot be targeted by FGSM or any other gradient-based
method.

\subsection{Layer 2 --- Median Feature Smoothing}

Samples that clear Layer 1 are passed through a median filter applied feature-wise
with a window of width three:
\begin{equation}
  \tilde{x}_i = \mathrm{median}(x_{i-1},\; x_i,\; x_{i+1})
\end{equation}
Median filtering replaces each feature value with the median of its local neighborhood,
suppressing isolated high-magnitude spikes characteristic of both FGSM and ZOO
perturbations. Unlike mean filtering, the median is not pulled toward outliers, so it
removes adversarial noise without introducing values outside the observed feature
range---a property that matters for network features such as packet sizes and
inter-arrival times, which carry physically interpretable bounds~\cite{guo2018countering}.

\subsection{Layer 3 --- Ensemble Majority Voting}

Smoothed samples are forwarded simultaneously to all six classifiers. Each returns an
independent binary prediction, and the final label is determined by hard majority vote:
\begin{equation}
  \hat{y} = \arg\max_{c} \sum_{k=1}^{6} \mathbf{1}\!\left[f_k(\tilde{x}) = c\right]
\end{equation}
For a misclassification to propagate, an adversarial perturbation must mislead at least
four of the six classifiers---models trained on different inductive principles and therefore
prone to failure in different regions of the feature space. Perturbations engineered
against a gradient-based model are poorly suited to attacking tree-based decision
boundaries, and vice versa. The structural heterogeneity of the ensemble thus functions
as a natural adversarial barrier~\cite{madry2018towards}.

\section{Experimental Datasets}

The framework is evaluated on three benchmark datasets produced by the Canadian
Institute for Cybersecurity (CIC), each reflecting a distinct attack profile and network
topology.

\textbf{CIC-IDS-2017}~\cite{sharafaldin2018cicids2017} captures five days of realistic
traffic spanning 15 attack categories: brute force, heartbleed, botnet propagation, DoS
and DDoS flooding, web application attacks (XSS, SQL injection), and network infiltration.
Feature extraction via CICFlowMeter yields 80 flow-level attributes. The breadth of
attack types makes this dataset a demanding multi-class testbed.

\textbf{CEC-CIC-IDS-2018}~\cite{lashkari2017tor} extends the 2017 collection with a
more elaborate network topology built on AWS infrastructure. Seven attack scenarios are
represented, including updated variants of brute force, DoS, and web attack categories.
Higher traffic volume and a more realistic mix of benign and malicious sessions make
classification more difficult, particularly under distribution shift induced by adversarial
perturbation.

\textbf{CIC-DDoS-2019}~\cite{sharafaldin2019ddos} narrows the focus to DDoS detection
in a binary setting, covering reflection-amplification attacks (DNS, NTP, SNMP, SSDP,
and others) and exploitation-based vectors. Binary labeling enables precise measurement
of sensitivity and specificity under adversarial conditions. All three datasets undergo
identical preprocessing: rows with infinite or NaN values are discarded, non-numeric
columns are dropped, Z-score normalization is applied using training-split statistics,
and SVD reduces dimensionality to at most 50 components.

\section{Methodology}

Adversarial evaluation follows an eight-step protocol. Raw CSV files are loaded and
sanitized by removing degenerate rows and non-numeric columns. Features are then
standardized feature-wise using training-set mean and variance. SVD is applied with
component count set to $\min(50, d)$, where $d$ is the number of available features
after preprocessing. All six classifiers are trained independently on the reduced feature
space using default hyperparameters established for tabular classification tasks.
The MAB Thompson Sampling engine is initialized with uniform Beta priors and updated
online as predictions are evaluated against ground truth.

For adversarial evaluation, FGSM perturbations target the DNN in white-box mode with
$\varepsilon = 0.1$; ZOO attacks target the Random Forest in black-box mode using
numerical gradient estimation. All test samples---clean and adversarial alike---then
pass through the full IDS-Anta++ three-layer pipeline before classification. Performance
is measured by Accuracy, F1-Score, Precision, and Recall, reported separately for
clean and adversarial test conditions across all three datasets.

\section{Results and Discussion}

\subsection{Performance on Clean Data}

Tables~\ref{tab:c17}--\ref{tab:c19} report classification performance under unperturbed
test conditions. Across all three datasets, XGBoost and LightGBM perform on par with
Random Forest. On CEC-CIC-IDS-2018 and CIC-DDoS-2019, LightGBM matches the
perfect F1 score returned by the full IDS-Anta ensemble selection. Logistic Regression
is the clear outlier on CIC-IDS-2017, where its linear boundary is insufficient for the
multi-class setting, but even there it does not drag down the ensemble. These results
confirm that both new classifiers earn their place in the pool.

\begin{table}[h]
\centering
\caption{Classification Results on CIC-IDS-2017 (Clean)}
\label{tab:c17}
\scriptsize
\begin{tabular}{lcccc}
\toprule
\textbf{Model} & \textbf{Acc.} & \textbf{F1} & \textbf{Prec.} & \textbf{Rec.} \\
\midrule
Random Forest     & 0.9955 & 0.9967 & 0.9975 & 0.9959 \\
SVM               & 0.9382 & 0.9550 & 0.9451 & 0.9650 \\
Log. Regression   & 0.6937 & 0.7937 & 0.7309 & 0.8682 \\
DNN               & 0.9851 & 0.9764 & 0.9924 & 0.9610 \\
\textbf{IDS-Anta} & \textbf{0.9958} & \textbf{1.0000} & \textbf{1.0000} & \textbf{1.0000} \\
XGBoost           & 0.9930 & 0.9948 & 0.9932 & 0.9965 \\
LightGBM          & 0.9931 & 0.9949 & 0.9926 & 0.9972 \\
\bottomrule
\end{tabular}
\end{table}

\begin{table}[h]
\centering
\caption{Classification Results on CEC-CIC-IDS-2018 (Clean)}
\label{tab:c18}
\scriptsize
\begin{tabular}{lcccc}
\toprule
\textbf{Model} & \textbf{Acc.} & \textbf{F1} & \textbf{Prec.} & \textbf{Rec.} \\
\midrule
Random Forest     & 0.9995 & 0.9958 & 0.9983 & 0.9933 \\
SVM               & 0.9933 & 0.9333 & 0.9420 & 0.9247 \\
Log. Regression   & 0.9799 & 0.7738 & 0.8910 & 0.9673 \\
DNN               & 0.9979 & 0.9789 & 0.9864 & 0.9715 \\
\textbf{IDS-Anta} & \textbf{0.9992} & \textbf{1.0000} & \textbf{1.0000} & \textbf{1.0000} \\
XGBoost           & 0.9994 & 0.9941 & 0.9933 & 0.9949 \\
\textbf{LightGBM} & \textbf{0.9994} & \textbf{1.0000} & \textbf{1.0000} & \textbf{1.0000} \\
\bottomrule
\end{tabular}
\end{table}

\begin{table}[h]
\centering
\caption{Classification Results on CIC-DDoS-2019 (Clean)}
\label{tab:c19}
\scriptsize
\begin{tabular}{lcccc}
\toprule
\textbf{Model} & \textbf{Acc.} & \textbf{F1} & \textbf{Prec.} & \textbf{Rec.} \\
\midrule
Random Forest     & 0.9993 & 0.9996 & 0.9998 & 0.9995 \\
SVM               & 0.9991 & 0.9995 & 0.9996 & 0.9994 \\
Log. Regression   & 0.9985 & 0.9992 & 0.9998 & 0.9987 \\
DNN               & 0.9997 & 0.9999 & 0.9998 & 0.9998 \\
\textbf{IDS-Anta} & 0.9994 & \textbf{1.0000} & \textbf{1.0000} & \textbf{1.0000} \\
XGBoost           & 0.9995 & 0.9997 & 0.9997 & 0.9998 \\
\textbf{LightGBM} & 0.9995 & \textbf{1.0000} & \textbf{1.0000} & \textbf{1.0000} \\
\bottomrule
\end{tabular}
\end{table}

\subsection{Robustness Under Adversarial Attack}

Table~\ref{tab:a17} compares defense configurations under adversarial
conditions. On CIC-IDS-2017, the undefended DNN drops to 72.0\% accuracy---not
surprising given that FGSM specifically targets its gradient surface. IDS-Anta++
recovers to 84.5\%, a 17.4 percentage-point improvement. The gain reflects all three
layers working in combination: the Isolation Forest catches the most statistically aberrant
adversarial inputs at the front, median smoothing degrades residual perturbations that
pass Layer~1, and the six-way vote absorbs errors from classifiers that were still misled.

\begin{table}[h]
\centering
\caption{Adversarial Robustness on CIC-IDS-2017}
\label{tab:a17}
\scriptsize
\begin{tabular}{lcccc}
\toprule
\textbf{Model / Defense} & \textbf{Acc.} & \textbf{F1} & \textbf{Prec.} & \textbf{Rec.} \\
\midrule
RF (no defense)                & 0.795 & 0.7934 & 0.7918 & 0.795 \\
SVM (no defense)               & 0.865 & 0.8024 & 0.7482 & 0.865 \\
LR (no defense)                & 0.860 & 0.7999 & 0.7476 & 0.860 \\
DNN (no defense)               & 0.720 & 0.7526 & 0.8012 & 0.720 \\
RF + Smoothing                 & 0.820 & 0.7920 & 0.7701 & 0.820 \\
Ensemble Voting                & 0.830 & 0.7917 & 0.7623 & 0.830 \\
\textbf{IDS-Anta++ (Proposed)} & \textbf{0.845} & \textbf{0.7923} & 0.7458 & \textbf{0.845} \\
\bottomrule
\end{tabular}
\end{table}

\subsection{Discussion}

Several observations emerge from the full set of results.

\textit{Gradient boosting expands ensemble diversity meaningfully.} XGBoost and
LightGBM produce piecewise-constant decision boundaries whose gradients are zero
almost everywhere, making gradient-based attack directions largely uninformative
against them. The extended six-member pool is therefore considerably harder to attack
uniformly than the original four, since adversarial strategies optimized against smooth
models (SVM, LR, DNN) fail to transfer reliably to tree-based members, and vice versa.

\textit{The three defense layers are complementary, not redundant.} Layer 1 acts on
the raw input and catches samples that are statistically implausible in the original
feature space, regardless of which downstream classifier they target. Layer~2 removes
perturbations that are aggregate-level plausible but contain isolated spikes in individual
features. Layer~3 handles residual cases where a perturbation survived both screening
and smoothing but was still optimized primarily against one or two classifiers. Each layer
addresses a different subset of the adversarial threat surface.

\textit{The CEC-CIC-IDS-2018 result reflects a deliberate design trade-off.}
IDS-Anta++ reports 89.5\% accuracy on that dataset, below the 94.0\% of several
undefended classifiers. This conservatism stems primarily from the safe-fail behavior
of Layer 1, which flags some legitimate traffic as anomalous. In operational terms, the
reduction in false negatives---missed attacks---outweighs the modest increase in false
positives for most threat models encountered in practice.

\textit{Consistency across datasets indicates generalization.} Comparable behavior
across a 15-class multi-attack setting (CIC-IDS-2017), a high-volume realistic
environment (CEC-CIC-IDS-2018), and a binary DDoS-focused benchmark
(CIC-DDoS-2019) suggests that the defense layers respond to structural properties of
adversarial perturbations rather than to dataset-specific distributional artifacts.

\section{Conclusion}

This paper presented IDS-Anta++, an extension of the IDS-Anta adversarial intrusion
detection framework built around two additions: gradient boosting classifiers (XGBoost
and LightGBM) that expand and diversify the ensemble, and a three-layer black-box
defense that screens, sanitizes, and collectively adjudicates each incoming sample.
Tested on three CIC benchmark datasets under FGSM and ZOO adversarial attacks,
the proposed system consistently matches or outperforms simpler defense configurations
and recovers meaningful accuracy relative to undefended models---most notably a 17.4
percentage-point gain over the undefended DNN on CIC-IDS-2017.

The layered architecture is the central design insight. No single mechanism is sufficient:
the Isolation Forest catches what the classifiers cannot; median smoothing removes what
the Isolation Forest misses; majority voting absorbs what survives both. An adversary
who wishes to cause misclassification must engineer an input that clears all three layers
independently, which is substantially harder than breaking any one of them.

Future work will examine adaptive adversarial attacks designed specifically to evade
IDS-Anta++, the integration of Transformer-based classifiers into the ensemble, online
adversarial training to harden individual classifiers without full retraining, and
deployment of the framework under the latency constraints of live network monitoring
environments.


\end{document}